\documentclass[pra,aps,twocolumn,floatfix,showpacs]{revtex4}

\usepackage[usenames,dvipsnames]{color}
\usepackage{graphicx}

\usepackage{amssymb}
\usepackage{amsmath}

\begin{document}

\title{Orthogonalization of partly unknown quantum states}

\author{M. Je\v{z}ek}
\affiliation{Department of Optics, Palack\'{y} University, 17. listopadu 1192/12, CZ-771 46 Olomouc, Czech Republic}

\author{M. Mi\v{c}uda}
\affiliation{Department of Optics, Palack\'{y} University, 17. listopadu 1192/12, CZ-771 46 Olomouc, Czech Republic}

\author{I. Straka}
\affiliation{Department of Optics, Palack\'{y} University, 17. listopadu 1192/12, CZ-771 46 Olomouc, Czech Republic}

\author{M. Mikov\'{a}}
\affiliation{Department of Optics, Palack\'{y} University, 17. listopadu 1192/12, CZ-771 46 Olomouc, Czech Republic}

\author{M. Du\v{s}ek}
\affiliation{Department of Optics, Palack\'{y} University, 17. listopadu 1192/12, CZ-771 46 Olomouc, Czech Republic}

\author{J. Fiur\'{a}\v{s}ek}
\affiliation{Department of Optics, Palack\'{y} University, 17. listopadu 1192/12, CZ-771 46 Olomouc, Czech Republic}

\begin{abstract}
A quantum analog of the fundamental classical NOT gate is a quantum gate that would transform
 any input qubit state onto an orthogonal state. Intriguingly, this universal NOT gate is forbidden by the laws
 of quantum physics. This striking phenomenon has far-reaching implications concerning quantum information processing 
 and encoding information about directions and reference frames into quantum states.  It also triggers the question under what conditions 
 the preparation of quantum states  orthogonal to input states becomes possible. Here we report on experimental 
 demonstration of orthogonalization of partly unknown single- and two-qubit quantum states. A state orthogonal 
 to an input state is conditionally prepared by quantum filtering, and the only required information about the input state is 
 a mean value of a single arbitrary operator. We show that perfect orthogonalization of partly unknown two-qubit entangled
  states can be performed by applying the quantum filter to one of the qubits only.
\end{abstract}


\maketitle 

\section{Introduction}

The laws of quantum physics impose fundamental limits on processing of information encoded into states of quantum systems. 
Our ability to extract information from a quantum register, represented by a sequence of spin 1/2 particles, depends on how 
the individual spins are oriented. Probably the most striking example consists of the higher efficiency of spin direction encoding 
into a pair of orthogonal spins compared to the parallel ones \cite{GisinPopescu99}. 
The different information capacities of orthogonal and parallel quantum states stem from their symmetry properties,
resulting in the impossibility to freely convert between these configurations. Explicitly, one cannot construct 
a perfect universal NOT gate that would map an arbitrary pure qubit state $|\psi\rangle$ onto an orthogonal 
qubit state $|\psi_\perp\rangle$, $\langle \psi_\perp|\psi\rangle=0$. An {\em imperfect} implementation of universal NOT gate is possible, although fundamentally limited by conservation laws \cite{vanEnk05}.
The average fidelity of the optimal approximate universal NOT gate reads $2/3$ \cite{Buzek99,Buzek00,DeMartini02} 
and it cannot be increased even if we allow for probabilistic operations \cite{Fiurasek04}. 
However, very recently, it has been shown by Vanner \emph{et al.} that the task of quantum state orthogonalization becomes feasible provided that we possess some a-priori information
about the input state \cite{Vanner}. In particular, it suffices to know a mean value $a= \langle \psi |A |\psi\rangle$ 
of some operator $A$. A state orthogonal to the
input state $|\psi\rangle$ can then be conditionally prepared by applying a quantum filter ${A}-a{{I}}$ to the input state,
\begin{equation}
|\psi_\perp\rangle \propto \left( {A}-  a {{I}}\right)|\psi\rangle,
\label{orthogonalization}
\end{equation}
where ${{I}}$ denotes the identity operator. It is simple to check that $\langle \psi_\perp|\psi\rangle=0$ as required. It follows from Eq. (\ref{orthogonalization}) that 
the success probability $p_\perp$ of the orthogonalization procedure can be expressed as
$p_\perp =(\langle A^\dagger A\rangle-|a|^2)/\lambda^2$, where $\lambda=\max_j|\Delta A_j|$ and $\Delta A_j$ denotes the singular values of $\Delta A=A -a {{I}} $.

Interestingly, the task of preparing a state orthogonal to a completely unknown input pure state $|\psi\rangle$ becomes easier 
with increasing Hilbert space dimension $d$. If the input states $|\psi\rangle$ are randomly chosen 
according to a uniform a-priori distribution induced by the Haar measure on $\mathrm{SU}(d)$, then the minimum achievable 
average overlap between input states $|\psi\rangle$ and output orthogonalized states reads  
\begin{equation}
F_{\perp}(d)=\frac{1}{d+1}.
\label{Fperp}
\end{equation}
For comparison, a hypothetical perfect orthogonalization device would achieve $F_\perp(d)=0$, where
by perfect orthogonalization we mean preparation of a state that is perfectly orthogonal 
to a given unknown quantum state. Formula (\ref{Fperp}) follows from the relation between 
average state fidelity $F$ and quantum process fidelity $F_{\chi}$ valid for an arbitrary deterministic quantum operation \cite{Horodecki99},
\begin{equation}
F=\frac{dF_{\chi}+1}{d+1}.
\end{equation}
Since $F_\chi\geq 0$ by definition, $F$ is minimized when $F_\chi=0$ and we obtain Eq. (\ref{Fperp}).
For unitary operations $U$, $F_\chi=|\mathrm{Tr}[U]|^2/d^2$, hence 
the minimum average overlap (\ref{Fperp}) can be achieved by any unitary operation that satisfies $\mathrm{Tr}[U]=0$.
Another option is to employ the universal quantum inverter \cite{Rungta00} that represents an extension of the
 approximate universal-NOT operation \cite{Buzek99} to qudits, $\mathcal{G}_{\mathrm{NOT}}(\rho)=\left(d{{I}} -\rho \right)/(d^2-1)$.
As discussed in Ref. \cite{Rungta00}, $\mathcal{G}_{\mathrm{NOT}}$ can be implemented by
a measure-and-prepare strategy. An isotropic measurement with POVM elements 
$|\varphi\rangle \langle \varphi| d\varphi$ is performed on the input state, and after obtaining a measurement outcome $|\varphi\rangle$
 an output state $({{I}} - |\varphi\rangle\langle \varphi|)/(d-1)$ is prepared. 

The above discussion suggests that the orthogonalization is most difficult for qubits while it becomes 
feasible for continuous variable states in  infinite dimensional Hilbert space. 
This can be intuitively understood by realizing that a continuous-variable state $|\psi\rangle$ 
can be coherently displaced by an arbitrary amount $\alpha$. The overlap between the displaced state $D(\alpha)|\psi\rangle$ 
and the input state $|\psi\rangle$ can be made arbitrarily small by choosing a sufficiently large displacement $\alpha$.

Here we focus on qubit systems and report on experimental perfect conditional orthogonalization 
of partly unknown pure single-qubit and two-qubit states encoded in polarization states of  photons 
generated by spontaneous parametric downconversion. The rest of the paper is organized as follows. In Section II we describe the orthogonalization procedure in detail
and in Section III we present our experimental setup. Experimental results for orthogonalization of partly unknown 
single-qubit states are discussed in Section IV. In Section V we consider orthogonalization of partly unknown two-qubit entangled states 
and we demonstrate that such states can be orthogonalized by a local operation where the quantum filter is applied to one of the qubits only. 
Finally, Section VI contains brief conclusions.

\begin{figure}[!b!]  
\centerline{\includegraphics[width=1.0\linewidth]{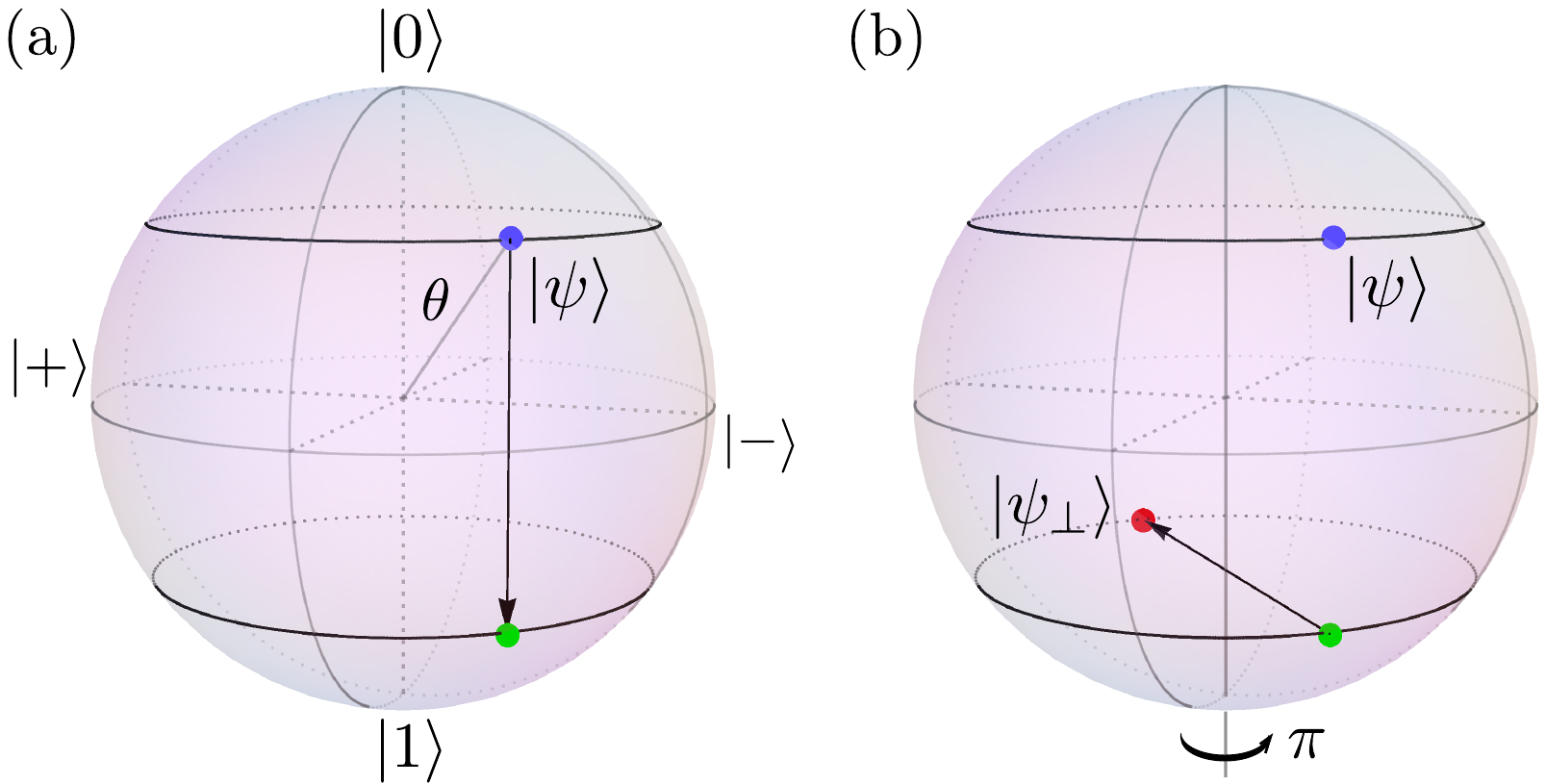}}
  \caption{(Color online) Orthogonalization of partly unknown single-qubit states. (a) Attenuation of amplitude of state $|0\rangle$. 
  (b) Unitary $\pi$ phase shift. For details, see text.}
\end{figure}

\section{Orthogonalization protocol}

In our study, the operator $A$ is chosen to be equal to the Pauli operator 
\begin{equation}
\sigma_Z=|0\rangle\langle 0|-|1\rangle\langle 1|.
\end{equation}
We parametrize the pure qubit states by spherical angles $\theta$ and $\phi$ on the Poincar\'{e} sphere, $|\psi\rangle=\cos \frac{\theta}{2}|0\rangle + e^{i\phi}\sin\frac{\theta}{2}|1\rangle$
and $|\psi_\perp\rangle=\sin \frac{\theta}{2}|0\rangle-e^{i\phi}\cos\frac{\theta}{2}|1\rangle$. 
Since $\langle\sigma_Z\rangle=\cos\theta$, the knowledge of $\langle \sigma_Z\rangle$ specifies the latitude on the Poincar\'{e} sphere, see Fig. 1. 
However, the state is still partly unknown, because $\phi$ can be arbitrary. Without loss of generality, we can assume that $\theta \leq \frac{\pi}{2}$, hence 
 $\langle \sigma_Z\rangle \geq 0$ and the  input state is located on the northern hemisphere of the Poincar\'{e} sphere. 
The quantum filter $Z \propto \sigma_Z- {{I}} \cos\theta $ producing a state orthogonal to $|\psi\rangle$ then reads, 
\begin{equation}
Z= \tan^2\frac{\theta}{2}\,|0\rangle \langle 0| -| 1\rangle \langle 1|. 
\label{Zdefinition}
\end{equation}
This operator is normalized such that the maximum of the absolute values of its eigenvalues is equal to $1$.

As illustrated in Fig. 1, the orthogonalization procedure can be divided into two steps. 
In the first step, the amplitude of the state $|0\rangle$ is attenuated by a factor of  $\tan^2(\theta/2)$. A circle on the Poincar\'{e} sphere, which is  specified  by  $\theta$, is transformed onto a similar circle symmetrically positioned 
with respect to the equator of the Poincar\'{e} sphere, $\theta'=\pi-\theta$.
In the second step, a unitary $\pi$ phase shift rotates this circle by $180^\circ$, $\phi'=\phi+\pi$, which maps all the states on the original circle onto orthogonal states. 
The orthogonalization procedure succeeds with a probability $p_\perp=\tan^2(\theta/2)$. $p_\perp$ is maximal
for states on the equator of Poincar\'{e} sphere ($\theta=\pi/2$), whose orthogonalization can be performed by a deterministic unitary $\pi$ phase shift \cite{Buzek99,Buzek00}. 
On the other hand, when we approach the limit $\theta=0$, then the amplitudes of the input states become highly unbalanced and heavy filtering
is required, which results in a small success probability. 

The orthogonalization procedure can be straightforwardly generalized to multipartite systems.
Consider a bipartite pure state  $|\Psi\rangle_{\mathrm{AB}}$. Suppose that we know a mean value of an operator $A$ acting on subsystem A,
$a =\langle \Psi| A_\mathrm{A}\otimes {{I}}_\mathrm{B}|\Psi\rangle$. Then  it holds that the state
\begin{equation}
|\Psi_\perp\rangle_{\mathrm{AB}} \propto \left(A -a {{I}}\right)_\mathrm{A} \otimes {{I}}_{\mathrm{B}} |\Psi\rangle_{\mathrm{AB}}
\end{equation}
is orthogonal to state $|\Psi\rangle_{\mathrm{AB}}$. The orthogonalization can thus be performed by local filtering operation on a single subsystem A.

\section{Experimental setup}

Our experimental setup is depicted in Fig. 2(a). Time correlated orthogonally polarized photon pairs with central wavelength of 810 nm are generated in the
process of type-II collinear spontaneous parametric downconversion in a 2~mm thick BBO crystal pumped by a CW laser diode with 75 mW of power and central wavelength of 405 nm \cite{Jezek11}.
The orthogonally polarized signal and idler photons are spatially separated on a polarizing beam splitter and coupled into single mode fibers. Detection of idler photon
heralds the presence of signal photon. The signal photon is released into free space and a desired input polarization state $|\psi\rangle$
is prepared with the help of a sequence of quarter- and half-wave plates. 

The filtering operation (\ref{Zdefinition}) was realized by a tunable polarization-dependent 
attenuator which consists of a pair of calcite beam displacers and half-wave plates \cite{Kwiat04,Kwiat05,Lemr11,Micuda12}.
The two beam displacers form an inherently stable Mach-Zehnder interferometer \cite{OBrien03}.
The first beam displacer introduces transversal spatial offset between vertically (V) and horizontally (H) polarized beams, and the half-wave plate HWP1 set at $45^\circ$ 
transforms the vertical polarization onto horizontal and vice versa. 
The polarization qubit $|\psi\rangle$ is thus converted into spatial qubit such that the states $|0\rangle$ and $|1\rangle$ correspond 
to the photon propagating in the upper and lower interferometer arms, respectively. The amplitude of photon propagating in the upper arm 
is selectively attenuated by rotating the half-wave plate HWP2. Rotation of HWP2 by angle $\vartheta$ transforms the initial horizontal polarization 
onto a linear polarization at angle $2\vartheta$. 
 The second beam displacer collects only the vertically polarized signal from the upper arm  while
the horizontally polarized signal is deflected and discarded. The amplitude attenuation factor of this device is thus given by  $\cos(2\vartheta)$.
  The output polarization state behind the second beam displacer was analyzed with the help of a detection block that 
  consists of a HWP and QWP followed by polarizing beam splitter and two single-photon detectors monitoring both output ports of the PBS.
  The unitary $\pi$ phase shift which is a part of the orthogonalization operation was in our implementation incorporated 
  into the setting of waveplates that are part of the detection block. 
  
\begin{figure}[!t!]  
  \centerline{\includegraphics[width=0.99\linewidth]{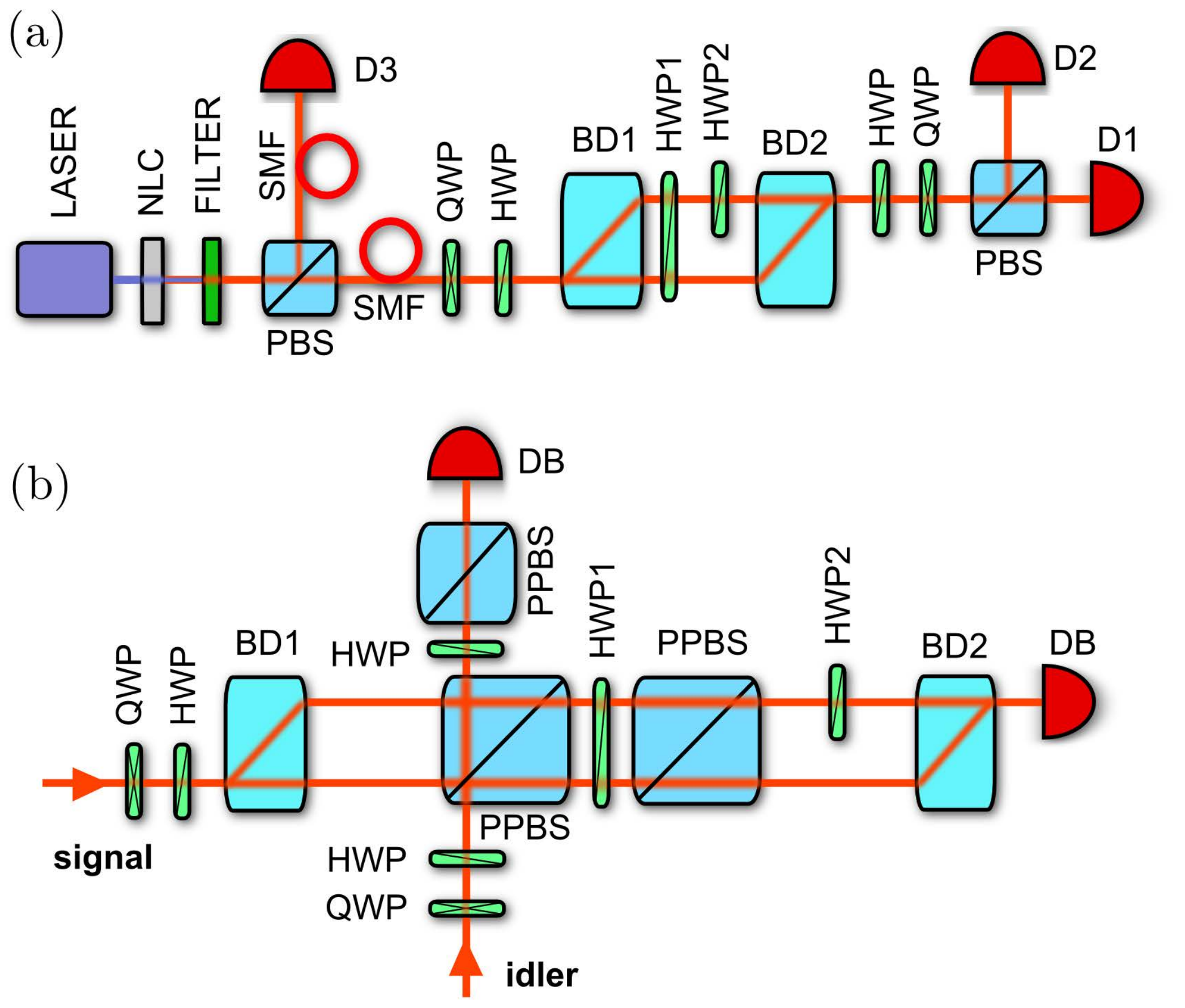}} 
\caption{(Color online) Experimental setup for orthogonalization of single-qubit (a) and two-qubit (b) states.  
  NLC---nonlinear crystal, SMF---single-mode fiber, 
  BD---calcite beam displacer, PPBS---partially polarizing beam splitter,
  PBS---polarizing beam splitter, HWP---half-wave plate, QWP---quarter-wave plate, 
  D---single-photon detector, DB---detection block consisting of a HWP, QWP, PBS
  and two single-photon detectors.}
\end{figure}

 \section{Single-qubit states} 
   
We have carried out measurements for 4 different values of $\theta$ and $\phi$, which represents in total $16$ different input single-qubit states $|\psi\rangle$. 
 For each input state, the HWP2 was first set to $\vartheta=0$ (no attenuation), and the input state was characterized by a 
 tomographically complete measurement consisting of a sequence of projective measurements in three mutually unbiased bases
$H/V$, $D/A$ and $R/L$ \cite{Wootters89,James01, Rehacek04,Altepeter05}. Here D and A denote the diagonally and anti-diagonally linearly polarized states, and
 R and L denote the right- and left-handed circularly polarized states. 
Then we set the attenuation according to the value of $\theta$ used in the state preparation procedure and performed 
quantum state tomography of the orthogonalized state. Finally, $\langle\sigma_Z\rangle$ was also
 estimated from projective measurement on the input state in the $H/V$ basis, the attenuation was set according to this measurement, 
 and a quantum state tomography of the output orthogonalized state was carried out. The states were reconstructed from the experimental data using the standard 
 maximum-likelihood estimation algorithm \cite{MaxLik}.

\begin{figure}[!b!]  
  \centerline{\includegraphics[width=\linewidth]{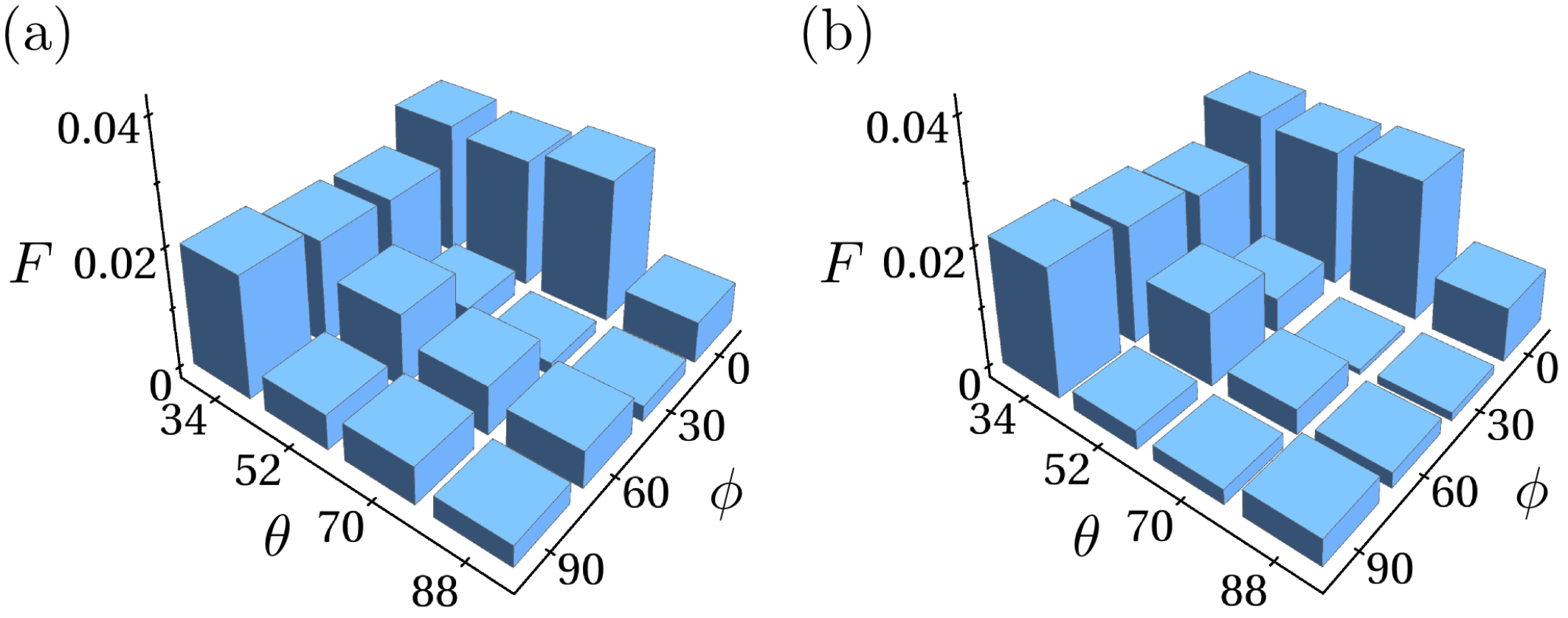}} 
  \caption{(Color online) Overlap $F$ between input and orthogonalized single-qubit states. The results are shown for the two approaches where $\langle\sigma_Z\rangle$ is determined either from the theoretical
  knowledge of the prepared input state (a) or from measurements on the input state in the $H/V$ basis (b).}
\end{figure}

The reconstructed input single-qubit states exhibited very high purity $\mathcal{P}=\mathrm{Tr}(\rho^2)$  
exceeding in all cases $0.992$. The orthogonalized states were slightly more mixed but the minimum observed purity was still as high as $0.986$.
We employ fidelity 
\begin{equation}
F= \left[\mathrm{Tr}\sqrt{\rho_1^{1/2}\rho_2 \rho_1^{1/2}}\right]^2
\end{equation}
to quantify the overlap of two mixed states $\rho_1$ and $\rho_2$. 
If $F=0$ then the two density matrices $\rho_1$ and $\rho_2$
have orthogonal supports. For single qubits it holds that $F=0$ if and only if both states are pure and orthogonal, $\rho_1=|\psi\rangle\langle\psi|$ and
$\rho_2=|\psi_\perp\rangle\langle\psi_\perp|$. The overlaps between input and orthogonalized states are plotted in Fig. 3. We can see that the overlap
is in all cases smaller than $0.0254$ which indicates good performance of the orthogonalization procedure. Note that the overlap is higher for smaller $\theta$.
A likely explanation of this feature is that small $\theta$ requires heavy filtering, as discussed above. In this case, any imperfection in 
setting of the attenuation factor can have a significant impact.

\begin{figure}[!t!]  
  \centerline{\includegraphics[width=0.9\linewidth]{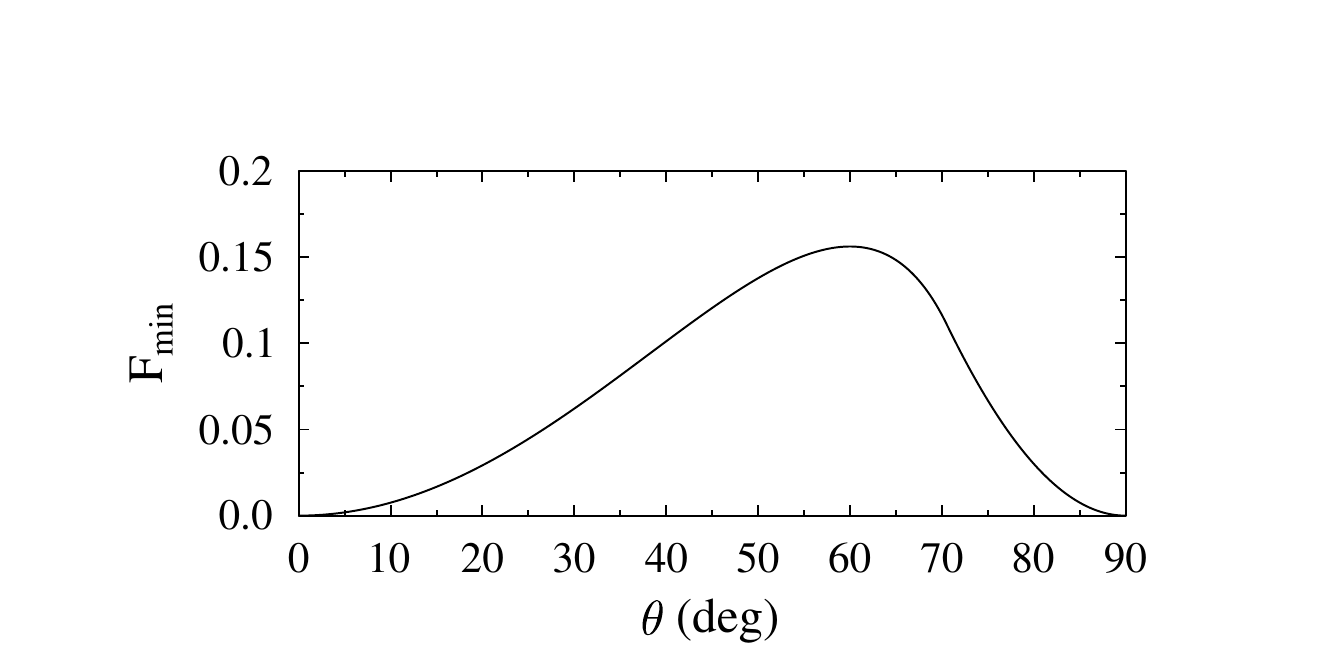}} 
  \caption{Minimum average overlap $F_{\mathrm{min}}$ between input and output single-qubit states 
  that is achievable by deterministic operations when $\theta$ is known.}
\end{figure}

\begin{figure}[!b!]  
  \centerline{\includegraphics[width=\linewidth]{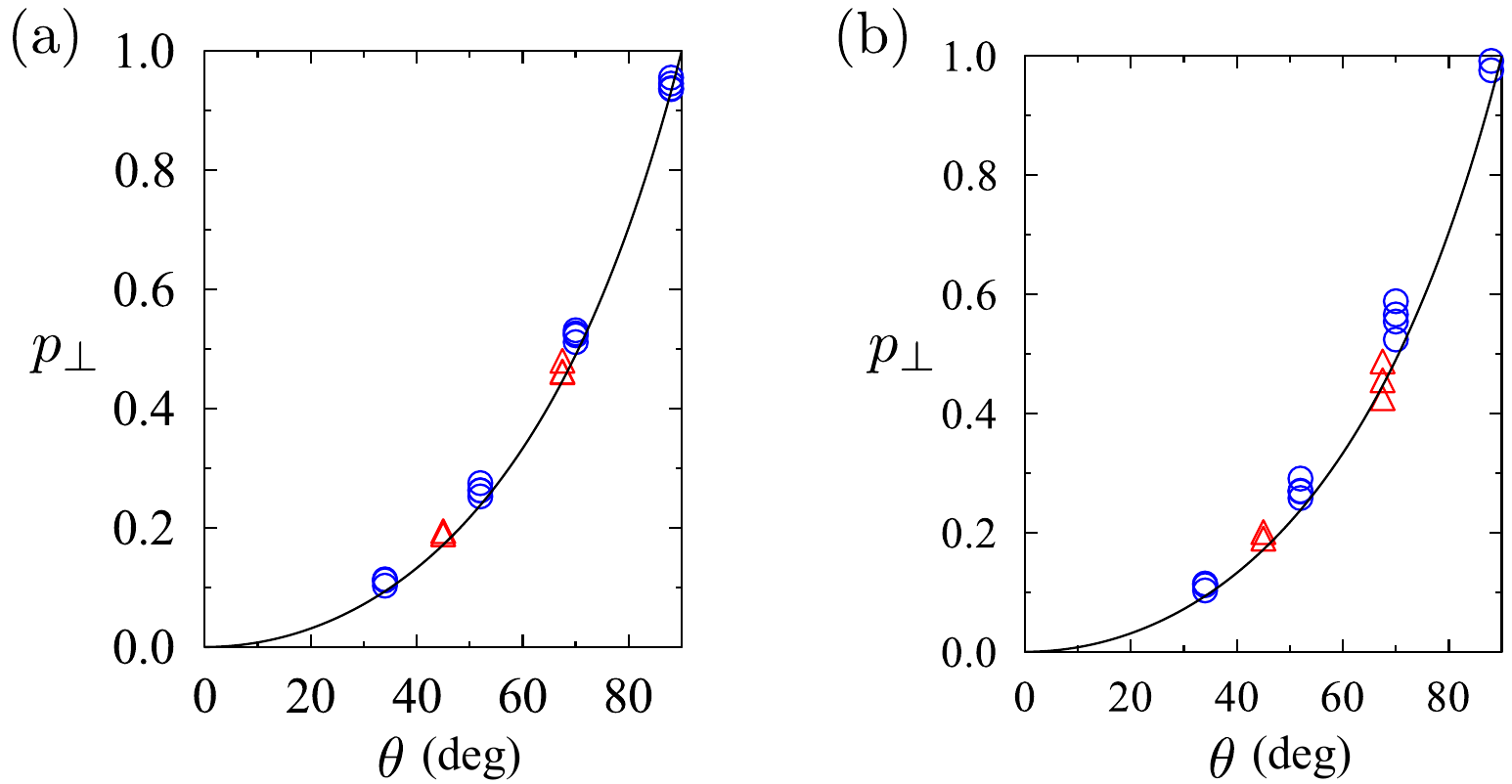}}
  \caption{(Color online) Success probability of orthogonalization is plotted as a function of $\theta$. The solid line represents theoretical dependence, and symbols indicate experimental results for single-qubit (blue circles)
  and two-qubit (red triangles) states. Results are shown for both methods of determination of $\langle \sigma_Z\rangle$,  as in  Fig.~3.}
\end{figure}

For comparison, we plot in Fig. 4 the minimum average overlap between input and output single-qubit states that is achievable by deterministic quantum operations when the input  states 
$|\psi\rangle$ have known fixed $\theta$. As shown in the Appendix, this minimum overlap is given by
\begin{equation}
F_{\mathrm{min}}=\left \{ 
\begin{array}{lcl}
\displaystyle \frac{1}{4}\sin^2\theta-\frac{\sin^6\frac{\theta}{2}}{\cos\theta}, & \quad & 0\leq \theta \leq \theta_{T}, \\[3mm]
\cos^2\theta, & & \theta_T < \theta \leq \frac{\pi}{2}, 
\end{array}
\right.
\label{Fmin}
\end{equation}
where $\theta_T=2\arcsin(1/\sqrt{3})$. We can see that $F_{\mathrm{\min}}$ vanishes only if the set of input states shrinks into a single state 
representing a pole of the Poincar\'{e} sphere, or if the states lie on the equator of the Poincar\'{e} sphere ($\theta=90^\circ$). 
 All experimentally determined overlaps plotted in Fig.~3, except those for $\theta=88^\circ$, 
lie well below $F_{\mathrm{min}}$. This confirms that the probabilistic orthogonalization outperforms the best deterministic strategy.

The success probability of conditional orthogonalization was determined as a ratio of the total number of measured coincidences for the orthogonalized 
and the input states, respectively, recorded over the time interval of $600$~s. The results are plotted in Fig. 5 and they agree well with the theoretical 
prediction. A higher vertical spread of data points corresponding to states with the same $\theta$ in Fig. 5(b) occurs because in this case the attenuation
was determined from measurements on input states. Therefore, the exact attenuation factors slightly varied among the states with identical
 $\theta$ but different $\phi$.

\section{Two-qubit entangled states}

We have also experimentally tested orthogonalization of partly unknown two-qubit entangled states by local single-qubit quantum filtration.
In our experiment, the two-photon entangled states were generated from input product states 
with the help of a linear optical quantum controlled-Z gate \cite{Ralph02,Hofmann02,Langford05,Kiesel05,Okamoto05}. 
As shown in Fig. 2(b), the two photons interfere on a partially polarizing beam splitter PPBS
with transmittances $T_V=1/3$ and $T_H=1$ for vertical and horizontal polarizations, respectively. This interference gives rise to a $\pi$ phase shift only if both
qubits are in logical state $|1\rangle$. The gate also includes two additional partially polarizing beam splitters which balance the amplitudes
and ensure unitarity of the gate. The gate operates in the coincidence basis which means that we have to post-select the events where
a single photon is detected in each output port of the gate \cite{Ralph02,Hofmann02,Langford05,Kiesel05,Okamoto05}. For technical reasons, the PPBS was placed 
inside the interferometer formed by the two beam displacers, see Fig. 2(b). The idler photon thus interferes with the signal photon only if the latter propagates through the lower interferometer arm. 
The setup is designed so that the signal photon propagating in the lower interferometer arm is vertically polarized, which ensures correct operation of the quantum CZ gate in this configuration.

\begin{table*}[!t!]
\caption{Overlap $F$ between the input (I) and orthogonalized (O) two-qubit states and purity $\mathcal{P}$ and entanglement of formation $E_f$ of the input and orthogonalized states. 
The data are presented for orthogonalization using the knowledge of $\langle \sigma_{Z1}\rangle$ from state preparation ($F$, $\mathcal{P}_O$, $E_{f,O}$) and
for orthogonalization where $\langle \sigma_{Z1}\rangle$ is determined from measurements on the first qubit ($F'$, $\mathcal{P}_O'$, $E_{f,O}'$).}
\begin{ruledtabular}
\begin{tabular}{cccccccccccc}
$\theta_1$ & $\phi_1$ & $\theta_2$ & $\phi_2$ & $F$ & $F'$ & $\mathcal{P}_I$ & $\mathcal{P}_O$ & $\mathcal{P}_O'$ &  $E_{f,I}$ & $E_{f,O}$ & $E_{f,O}'$ \\
$45^\circ$ & $0^\circ$ & $90^\circ$ & $0^\circ$ & 0.040 & 0.044 &  0.964 &  0.890 & 0.909 & 0.547 & 0.612   & 0.622 \\
$67.5^\circ$ & $0^\circ$ & $90^\circ$ & $0^\circ$ & 0.031 & 0.037 &  0.961 & 0.891 & 0.907 & 0.819 & 0.807    & 0.781\\
$45^\circ$ & $0^\circ$ & $45^\circ$ & $0^\circ$ & 0.021 & 0.029 & 0.936 &   0.944  & 0.942 &  0.286 & 0.334  & 0.361 \\
$67.5^\circ$ & $0^\circ$ & $45^\circ$ & $0^\circ$ & 0.008 & 0.008  & 0.975 & 0.952  & 0.941 & 0.523 &  0.482 & 0.496 \\
$67.5^\circ$ & $90^\circ$ & $45^\circ$ & $90^\circ$ & 0.041 & 0.035 & 0.971  & 0.946 & 0.935 & 0.497 & 0.518 &  0.468
 \end{tabular}
\end{ruledtabular}
\end{table*}

Let $\theta_1$, $\phi_1$ and $\theta_2$, $\phi_2$ denote the parameters of the input single-qubit states of signal ($|\psi_1\rangle$) and idler ($|\psi_2\rangle$) photon, respectively. 
The quantum CZ gate is diagonal in the computational basis, $U_{CZ}|jk\rangle=(-1)^{jk}|jk\rangle$, and for input product state $|\psi_1\rangle|\psi_2\rangle$ we obtain
\begin{equation}
|\Psi\rangle=U_{CZ}|\psi_1\rangle|\psi_2\rangle=\cos \frac{\theta_1}{2}|0\rangle|\psi^{+}\rangle+ e^{i\phi_1}\sin \frac{\theta_1}{2}|1\rangle|\psi^{-}\rangle, 
\end{equation}
where $|\psi^{\pm}\rangle= \cos \frac{\theta_2}{2}|0\rangle \pm e^{i\phi_2}\sin \frac{\theta_2}{2}|1\rangle$. 
Since $\langle\sigma_{Z1}\rangle=\cos\theta_1$ the amount of filtering required for orthogonalization depends only on $\theta_1$. 

After preparation of the input entangled state $|\Psi\rangle$, filtering operation (\ref{Zdefinition}) can be applied to the first qubit by rotating HWP2.
Polarization states of both photons are then measured with the help of two detection blocks DB identical to that shown in Fig. 2(a) and coincidences between clicks of detectors in the two blocks
are counted. We have performed full tomographic reconstruction of the input entangled states
$|\Psi\rangle$ as well as of the orthogonalized states. Like for single-qubit states, we have first used the theoretical 
value of $\langle \sigma_{Z1}\rangle$ known from state preparation and then we have
also used the value determined from measurements on the first qubit in the $H/V$ basis.

 The experimental results are summarized in Table I. 
 A successful orthogonalization is indicated by low overlaps between input and orthogonalized states. The purities of the orthogonalized states 
 are generally lower than the purity of the input state. This occurs because the filtration effectively enhances the
 terms sensitive to the visibility of two-photon interference on the central PPBS. In our experiment, we have measured visibility $\mathcal{V}=0.94$. The success probability of orthogonalization
 is plotted in Fig. 5 (red triangles) and the results agree well with the theory. 
 
 We have also determined entanglement of formation of the input and orthogonalized states \cite{Wootters98}. The values are listed in Table I
 and they confirm that the states are highly entangled. Due to various experimental imperfections, the observed $E_{f,I}$ is slightly lower than the theoretically predicted entanglement 
 of pure two-qubit state $|\Psi\rangle$ which can be expressed as $S_E=-x\log_2 x -(1-x)\log_2(1-x)$, where $x=\frac{1}{2}\left(1+\sqrt{1-\sin^2\theta_1\sin^2\theta_2}\right)$. 
This latter formula also indicates that in the present case the orthogonalization should preserve the amount of entanglement, because $S_E$ is invariant with respect 
to the transformation $\theta_1\rightarrow \pi-\theta_1$. The differences between the measured $E_{f,I}$ and $E_{f,O}$ are indeed rather small and can be attributed to the fact that 
the experimentally generated input states are not entirely pure and slightly differ from the theoretical states $|\Psi\rangle$.

\section{Conclusions}

In summary, we have experimentally demonstrated orthogonalization of partly unknown single-qubit and two-qubit states by quantum filtering.
Our experimental data clearly show that if we possess some partial prior information about the state that should be orthogonalized, then conditional orthogonalization
significantly outperforms the best deterministic procedure. Remarkably, bipartite entangled states can be orthogonalized by a local strategy where the quantum filter is applied 
just to one of the qubits and no information about the state of the other qubit is necessary. The conditional orthogonalization represents an intriguing addition 
to the toolbox of probabilistic protocols such as unambiguous quantum 
state discrimination \cite{IDP87a,IDP87b,IDP87c}, probabilistic quantum cloning \cite{Duan98,Muller12}, 
and quantum metrology assisted with abstention \cite{Gendra13}. We anticipate applications of the orthogonalization procedure 
in quantum information processing and quantum state engineering.

\acknowledgments
This work was supported by the Czech Science Foundation (Project No. 13-20319S) and by Palack\'{y} University (Project No. PrF-2013-008).

\appendix*
\section{Deterministic orthogonalization of single-qubit states with prior information}

Let us consider single-qubit input states  
\begin{equation}
|\psi\rangle=\cos\frac{\theta}{2}|0\rangle+e^{i\phi}\sin\frac{\theta}{2}|1\rangle
\label{psiin}
\end{equation}
with known  $\langle \sigma_Z\rangle$, i.e. with known fixed $\theta$.
Here we derive the minimum average overlap between input states (\ref{psiin}) and output states $\mathcal{E}(|\psi\rangle\langle \psi|)$, which is 
achievable by deterministic quantum operations $\mathcal{E}$, 
 i.e. by trace-preserving completely positive maps. According to the  Choi-Jamiolkowski isomorphism \cite{Jamiolkowski72,Choi75}, 
 any trace preserving completely positive map is isomorphic to a positive semidefinite operator $\chi$ on the tensor product 
 of the Hilbert spaces of the input and output states. Given an input state $\rho_{\mathrm{in}}$
 the corresponding output state can be calculated as 
 $\rho_{\mathrm{out}}=\mathrm{Tr}_{\mathrm{in}}[\rho_{\mathrm{in}}^T\otimes{{I}}_{\mathrm{out}} \,\chi]$, where $T$ stands for a transposition
 in a fixed basis. The trace preservation condition can be expressed as 
 \begin{equation}
 \mathrm{Tr}_{\mathrm{out}}[\chi]={{I}}_{\mathrm{in}}.
 \label{chitrace}
 \end{equation}
 
Assuming homogeneous prior distribution of angle $\phi$, the average overlap
between input and output states achieved by quantum operation $\chi$ can be expressed as
 \begin{equation}
 F_\theta=\frac{1}{2\pi}\int_{0}^{2\pi} \mathrm{Tr}[\psi^T\otimes \psi \, \chi] d\phi,
 \label{Fintegral}
 \end{equation}
 where $\psi=|\psi\rangle\langle \psi|$. This  integral can easily be evaluated, and we get
\begin{equation}
F_\theta=\mathrm{Tr}[R_\theta\chi],
\end{equation}
where the operator $R_\theta$ is given by
\begin{eqnarray}
R_\theta&=& c^4 |00\rangle\langle 00|+c^2s^2(|01\rangle\langle 01|+|10\rangle\langle 10|) +s^4 |11\rangle\langle 11| \nonumber \\
& &+c^2s^2(|00\rangle \langle 11|+|11\rangle \langle 00|).
\end{eqnarray}
Here we introduced abbreviations 
\begin{equation}
c=\cos \frac{\theta}{2}, \qquad s=\sin\frac{\theta}{2}.
\end{equation}
The optimal trace-preserving quantum operation that minimizes the average overlap $F_\theta$ can be determined by solving 
a semidefinite program \cite{Vandenberghe96,Audenaert02}. We shall first present the resulting operation and then prove its optimality.  For the sake of simplicity 
we shall restrict ourselves to the northern hemisphere of the Poincar\'{e} sphere, $\theta \leq \pi/2$. Then, the optimal $\chi$ can be expressed as
\begin{equation}
\chi_{\mathrm{opt}}=(a |00\rangle- |11\rangle)(a\langle 00|-\langle 11|)+(1-a^2)|01\rangle\langle 01|,
\end{equation}
where the parameter $a$ depends on $\theta$ as follows,
\begin{equation}
a= \left \{ 
\begin{array}{lcl}
\displaystyle \frac{\sin^2\!\frac{\theta}{2}}{\cos\theta}, & \quad & 0\leq \theta \leq \theta_{T}, \\[3mm]
1, & & \theta_T < \theta \leq \frac{\pi}{2}.
\end{array}
\right.
\end{equation}
The threshold angle  $\theta_T=2\arcsin(1/\sqrt{3})$ is determined by the condition $2s^2=c^2=2/3$.
The average overlap achieved by the optimal operation $\chi_{\mathrm{opt}}$ reads
\begin{equation}
F_{\mathrm{min}}=\left \{ 
\begin{array}{lcl}
\displaystyle \frac{1}{4}\sin^2\theta-\frac{\sin^6\frac{\theta}{2}}{\cos\theta}, & \quad & 0\leq \theta \leq \theta_{T}, \\[3mm]
\cos^2\theta, & & \theta_T < \theta \leq \frac{\pi}{2}. 
\end{array}
\right.
\end{equation}

If $s^2>1/3$, then $a=1$ and $\chi_{\mathrm{opt}}$ represents a unitary $\pi$ rotation about the $z$ axis, $|\psi\rangle \rightarrow \sigma_Z |\psi\rangle$,
which perfectly orthogonalizes states lying on the equator of the Poincar\'{e} sphere \cite{Buzek99}. If we get close enough
to the north pole of the Poincar\'{e} sphere such that $\theta < \theta_T$, then the optimal operation becomes a sequence of a unitary transformation $\sigma_Z$ 
and an amplitude damping channel, where the state $|0\rangle$ decays into state $|1\rangle$ with probability $1-a^2$.

To prove the optimality of $\chi_{\mathrm{opt}}$, we first define an operator $\lambda=\mathrm{Tr}_{\mathrm{out}}[R_\theta\chi]$. We get
\begin{equation}
\lambda= \left\{ \begin{array}{ll}
\displaystyle \frac{1}{4}\sin^2\theta|0\rangle\langle 0|-\frac{\sin^6\frac{\theta}{2}}{\cos\theta}|1\rangle\langle 1|, & 0 \leq \theta \leq \theta_T,\\[3mm]
\displaystyle  \cos\theta\left(\cos^2\frac{\theta}{2}|0\rangle\langle 0| -\sin^2\frac{\theta}{2} |1\rangle \langle 1|\right), & \theta_T < \theta \leq \frac{\pi}{2}.
\end{array}
\right.
\end{equation}
It holds by definition that $F_{\mathrm{min}}=\mathrm{Tr}[\lambda]$. We now prove that the operator 
\begin{equation}
M= R_\theta-\lambda\otimes {{I}}
\end{equation}
is positive semidefinite, $M\geq 0$. This implies that $F_{\mathrm{min}}$ is the minimum achievable over all deterministic quantum operations $\chi$. Indeed, since $\chi \geq 0$, we have
$\mathrm{Tr}[M\chi] \geq 0$, which yields
\begin{equation}
\mathrm{Tr}[R_\theta\chi] \geq \mathrm{Tr}[\lambda \otimes {{I}}\, \chi]= \mathrm{Tr}[\lambda]=F_{\mathrm{min}}.
\end{equation}
Here we used the trace-preservation condition (\ref{chitrace}). 
If $0 \leq \theta \leq \theta_T$, then the eigenvalues of $M$ read
\begin{equation}
\begin{array}{lcl}
\displaystyle m_1=0, & \quad & \displaystyle m_3=c^2(c^2-s^2)+\frac{c^2s^4}{c^2-s^2}, \\[3mm]
\displaystyle  m_2=0, & & \displaystyle  m_4=c^2s^2+\frac{s^6}{c^2-s^2}. \\[2mm]
~~
\end{array}
\end{equation}
Since $c^2 >s^2$ for all $0\leq \theta \leq \theta_T$, all eigenvalues $m_j$ are non-negative. If $\theta_T <\theta \leq \pi/2$, then
the eigenvalues read
\begin{equation}
\begin{array}{lcl}
\displaystyle m_1=0, & \quad &  m_3=c^2(2s^2-c^2), \\[3mm]
\displaystyle  m_2=2c^2s^2, & & m_4=s^2(2c^2-s^2).
\end{array}
\label{meigtwo}
\end{equation}
In this case $2s^2>c^2\geq s^2$ (see the definition of $\theta_T$ above), which ensures that all eigenvalues (\ref{meigtwo}) ar also non-negative.
This concludes the proof of the optimality of $\chi_{\mathrm{opt}}$. Due to symmetry, the optimal operation for $\theta> \pi/2$ can be obtained from the optimal operation
for $\pi-\theta$ by bit flips on both input and output qubits, $|0\rangle \rightarrow |1\rangle$, $|1\rangle \rightarrow |0\rangle$.

\end{document}